\documentclass[aps,pra,letterpaper,10pt,twocolumn,nofootinbib,superscriptaddress]{revtex4-2}%superscriptaddress
\usepackage[colorlinks=true,allcolors=blue]{hyperref}
\usepackage{amssymb}
 \usepackage{mathptmx}
\usepackage{amsmath}
\usepackage{graphicx}
\usepackage{subfigure}
\usepackage{amsfonts}
\usepackage{xcolor}
\usepackage{epsfig}
\usepackage{color}
\usepackage{bbm}
\usepackage{bm}
\usepackage{tabularx}
\usepackage{multirow}
\usepackage{mathtools}
\usepackage{xfrac}
\usepackage{scalerel}
\DeclareMathAlphabet{\mathcal}{OMS}{cmsy}{m}{n}

\DeclareMathOperator{\csch}{csch}

\newcommand{\ket}[1]{\vert{#1}\rangle} 
\newcommand{\bra}[1]{\langle{#1}\vert} 
 
\newcommand{\proj}[1]{\ket{#1}\!\bra{#1}}
\newcommand{\trans}[2]{\ket{#1}\!\bra{#2}}
\newcommand{\mean}[1]{\langle #1 \rangle}
\newcommand{\one}{\openone}
\newcommand{\half}{\tfrac{1}{2}}
\newcommand{\Nbos}{\overline{n}}

\newcommand{\tr}[2]{\mathrm{Tr}_{#1}#2}
\newcommand{\calC}{\mathcal{C}}
\newcommand{\calK}{\mathcal{K}}
\newcommand{\calW}{\mathcal{W}}
\newcommand{\supD}{\widehat{\rm D}}
\newcommand{\supL}{\widehat{\rm L}}
\newcommand{\supP}{\widehat{\rm P}}
\newcommand{\supQ}{\widehat{\rm Q}}
\newcommand{\supT}{\widehat{\rm T}}
\newcommand\Dag{\stretchrel*{\dag}{X}}

\begin{document}
\title{Non-classicality of squeezed non-Markovian processes}

\author{Mehdi Abdi}
\email{mehabdi@gmail.com}
\affiliation{Wilczek Quantum Center, School of Physics and Astronomy,
Shanghai Jiao Tong University, Shanghai 200240, China}
\affiliation{Department of Physics, Isfahan University of Technology, Isfahan 84156-83111, Iran}

\author{Moslem Zarei}
\affiliation{Department of Physics, Isfahan University of Technology, Isfahan 84156-83111, Iran}

\date{\today}
\begin{abstract}
We study nonclassical effects in the dynamics of an open quantum system.
The model involves a harmonic oscillator coupled to a reservoir of non-interacting harmonic oscillators.
Different system-bath interaction schemes as well as reservoir states are considered.
Particularly, the squeezed reservoirs coupled to the system through single and two quanta exchange processes are put in the spotlight.
We investigate the quantumness conveyed to the system through the bath by computing a nonclassicality measure for different bath properties and when the memory effects are appreciable.
The measure of nonclassicality is calculated for projective measurements both in the number state basis and a basis formed by a set of coherent states.
Our results show that in both bases the measure exhibits characteristic features for each bath state and the form of its interaction with the system.
Some of those features are independent from the measurement scheme (number or coherent), and thus, emergent from the bath and its interaction with the probe system.
This allows for fingerprinting and identifying the environmental effects by tracking a given probe with appropriate measurements.
Hence, may prove useful for distinguishing different sources of decoherence.
\end{abstract}

\maketitle

%
%
%----------INTRO----------%
\section{Introduction}
The quantumness of environmental effects in an open quantum system is a fundamental question in physics that is increasingly attracting the attention of quantum physicists~\cite{Budini2018, Knee2018, Munoz2020, Seif2022}.
One of the main questions in this concept is whether the environmental noise imposed on a quantum system is necessarily a quantum effect.
That is, the long trusted open quantum system belief that the decoherence stems from the entanglement of the bath degrees of freedom to the system is being challenged, as in some cases it can be modeled by purely classical effects~\cite{Chen2018, Gu2019}.
Therefore, the answer to the question that when a system-reservoir interaction is nonclassical and how one can trace it seems crucial.
Meanwhile, the effect of bath memory and its effect on the quantum behavior of the system of interest still needs some theoretical clarification and experimental confirmation.
Hence, the nonclassicality of Markov and non-Markovian reservoirs have been discussed and investigated in several recent works~\cite{Ban2007, Smirne2018}.
Mostly, bringing up the conclusion that a non-Markovian bath is \textit{more} quantum mechanical~\cite{Li2021}.

A common and widely accepted scheme for understanding and quantifying the physical systems and their properties is determining the correlations among its components.
Among them are the multitime correlations which can reveal various aspects of a system and because of that have been widely exploited in the experiments~\cite{Gardiner2010}.
Indeed when nonclassical effects are concerned, multitime measurements are believed as the only reliable probe for revealing the quantum correlations formed in a system as the classical fluctuations can also create quantum coherence, see e.g.~\cite{Trapani2015}.
Hence, the multitime correlations have been used in various schemes to evade such misunderstandings.
This includes the famous Leggett-Garg inequality that targets macrorealism of quantum measurements~\cite{Leggett1985}.
This inequality has been applied and proposed for the study of quantum features in a variety of systems; from the harmonic oscillators~\cite{Asadian2014, Bose2018} to two-level systems~\cite{Ali2017, Li2021}.
The reliability of the multitime measurements have even led to going beyond the dichotomic variables and resulted in introducing witnesses and measures for revealing the quantum essence of dynamics when multiple measurement outcomes are possible~\cite{Li2012}.
Measures of this type take violation of the Kolmogorov conditions as the test for nonclassical behaviors, which in turn, trials invasiveness of the measurements.
Despite its complications in the non-Markovian processes, defining such a measure is possible~\cite{Milz2020}.

The non-Markovian reservoirs are believed to present enhanced quantumness when the same type of interaction with the surrounding environment comes into the study~\cite{Budini2018, Li2021, Gholipour2020, Zhang2020}.
However, one could also ask about the effects that emerge due to the different forms of the system-bath interaction.
More specifically, the single- and multi-quanta interactions between a system and its reservoir have been envisaged through different systems~\cite{Gilles1993} or even engineered~\cite{Leghtas2015}.
Hence, highlighting their differences and putting forth a method for their identification can be useful for revealing the linearity and nonlinearity of the system-reservoir interactions in open quantum processes.
Along the same path one may ask for intensity of the nonclassical traces that either of these processes leave in the system.
Furthermore, impacts of the reservoir state and its specifications on the quantum dynamics of the system interacting with needs to be answered.
It is the main goal of the current study to address these questions to some extend.

Here, we study the nonclassicality built-up in a simple quantum system, a harmonic oscillator, coupled to a reservoir of bosonic modes.
We investigate the effect of bath properties; state of its modes as well as the way they couple to the system, and compute the quantumness measure originally introduced by Li \textit{et al} \cite{Li2012} and later formalized and generalized to non-Markovian processes in Ref.~\cite{Milz2020}.
The measure depends on the chosen projective measurement basis.
Therefore, given the continuous variable nature of the system of our study, we consider measurements in both Fock basis and the phase space.
For the latter, to elude the divergencies in the measure---resulting from the over-completeness of the coherent state---we propose to granulate the phase space and consider a finite set of coherent states as the basis that optimally cover the area of interest in the phase space.
The system dynamics is modeled and studied through the quantum optical master equation.
By considering both Markovian and non-Markovian cases we come to the conclusion that a Markovian bath, either with bilinear or nonlinear coupling, leaves negligibly small traces of nonclassicality as long as its components assume separable states.
That is, when the bath modes are not in an bipartite or multipartite entangled state~\cite{Tanimura2020, Abdi2021} the nonclassicality measure remains vanishing.
In the bilinear and non-Markovian case the witness is only significant when the bath is in a squeezed state.
We reason that this indeed roots back to the nature of the associated dissipators that can evolve an initially diagonal density matrix into a state with off-diagonal elements, and thus, building-up a time-correlated quantum coherence in the system.
Moreover, by investigating different frequency dependencies for the squeezing parameter of the bath, it is found that the measure can be employed for characterizing a bath through the interacting probe system.
With a cross check of the coherent basis measurement results as well as the various initial states of the system one may fingerprint the bath for revealing its microscopic properties.
More specifically, the microscopic form of the interaction can be tracked.
%The conveyed quantumness to the system is significantly enhanced when the coupling is in the form of two photon process where a single excitation of the system is exchanged with two excitations of the reservoir modes.

The rest of paper is organized as follows:
In the next section we discuss about the detection schemes and the nonclassicality measure employed in this work.
In Sec.~\ref{sec:model} the formulation of the microscopic model is provided.
Then we present our numerical results for the bilinear (Sec.~\ref{sec:bilinear}) and the nonlinear (Sec.~\ref{sec:nonlinear}) system-bath interactions.
The concluding remarks are given in Sec.~\ref{sec:conclusion}.

%
%
%----------QUANTUMNESS----------%
\section{Nonclassicality}\label{sec:measure}
We first discuss the detection schemes.
Correlation functions are key features in characterizing the nature of physical systems the noises affecting them.
Even though single-time measurements can reveal the presence of decoherence, they cannot determine its deeper nature, and specifically, whether it is a classical process.
Therefore, performing multi-time measurements are proposed to find out the underlying essence, among which two-time measurements are the most convenient ones.
A scheme based on such measurements can give a measure for testing the nonclassicality of an evolution.
In this scheme the system, as the probe, is initially prepared in a diagonal state in the computational basis of interest.
For simplicity we assume that the system is prepared in the eigenstate $\ket{x_0}$ at time $t=t_0$.
A projective measurement is then performed at $t=t_1$ and the outcome $x_1$ is obtained with the probability $P(x_1,t_1)$.
The system is then left to evolve until time $t=t_2$ when the second measurement is performed.
The outcome $x_2$ and the probability of finding it may depend on the first measurement outcome and probability.
Hence, the joint probability distribution $P(x_2,t_2;x_1,t_1)=P(x_2,t_2|x_1,t_1)P(x_1,t_1)$ plays a pivotal role in describing dynamics of a system.
Given the noninvasiveness essence of the measurements in classical physics $P(x_2,t_2;x_1,t_1)$ satisfies the Kolmogorov consistency condition: $\sum_{x_1}P(x_2,t_2;x_1,t_1)=P(x_2,t_2)$, where the sum on the left hand side is over all possible measurement outcomes.
With these, the following witness can be employed to find out and \textit{measure} the nonclassical nature of a process
\begin{equation}
\calW_Q = \sum_{x_2}\Big| P(x_2) -\sum_{x_1} P(x_2;x_1) \Big|,
\label{witness}
\end{equation}
where we have removed the explicit indication of the time arguments, implying that the outcome $x_k$ refers to the measurement time $t_k$.
This was primarily introduced and pointed out in Ref.~\cite{Li2012} as a witness and later formalized as a measure and generalized to non-Markovian reservoirs in Ref.~\cite{Milz2020}.

Here, we choose both the number and the coherent states as the computational basis and investigate the possibilities for achieving $\calW_Q >0$ for varieties of system and bath states and interactions.
Therefore, the trivial initial state depending on employed scenario is either a number state $\ket{n_0}$ or a coherent state $\ket{\alpha_0}$.
Note that one must consider the over-completeness of the coherent states.
Measurement in the coherent basis can be performed via heterodyne detection which gives two outcomes, the real and imaginary parts of the coherent amplitude with the probability given by the Husimi $Q$-function~\cite{Wiseman1996, Schleich2001}.
And, in principle, it gives a continuum of outcomes.
This calls for modifying the witness into
$\calW_Q = \int\!d^2\!\alpha_2\left| P(\alpha_2,\alpha_2^*,t_2) -\int\!d^2\!\alpha_1 P(\alpha_2,\alpha_2^*,t_2;\alpha_1,\alpha_1^*,t_1) \right|$, where now $P$ is a probability distribution function.
Nevertheless, to cover the continuum of the phase space one requires to perform infinite number of heterodyne measurements, which is not practical.
One, thus, granulates the phase space such that the overlaps are manageable.
In the following study we have considered different granulations and choose an optimal set of coherent states that cover the phase space properly and yet they have little overlap.

%
%
%----------MODEL----------%
\section{Model}\label{sec:model}
The model considered in this work is a harmonic oscillator coupled to a reservoir of harmonic oscillators.
This limits the nonlinearity down to the interaction between the reservoir modes and the system, and thus allows for an easier tracking of the nonlinear effects.
We first derive a master equation that describes the effective dynamics of the system by tracing out the reservoir degrees of freedom.
The general form of the Hamiltonian reads ($\hbar=1$)
\begin{equation}
H = \widetilde\omega a^\dag a^{} +\sum_{\bf k} \Omega_{\bf k}b_{\bf k}^\dag b_{\bf k}^{} + q \sum_{\mathbf{k}} g_{\bf k} Q_{\bf k},
\label{hamiltonian}%
\end{equation}%
where the system bare frequency is $\widetilde\omega$, and $\Omega_{\bf k}$ are the frequencies of the reservoir.
The system and bath annihilation and creation operators follow the bosonic commutation relations $[a,a^{\Dag}]=1$, $[b_{\mathbf{k}'}^{},b_{\bf k}^{\Dag}]=\delta_{\mathbf{k}\mathbf{k}'}$, and all other commutators are vanishing.
Here, $q$ ($Q_{\bf k}$) is the system (bath) interaction operator that we shall take it in the form of $q \propto a^n+a^{\Dag n}$ ($Q_{\bf k} \propto b_{\bf k}^{m} +b_{\bf k}^{\Dag m}$) to include the nonlinear interactions.
We put our focus on the most relevant cases where one deals with $n=1,2$ and $m=1,2$.
$g_{\bf k}$ the coupling rate of the reservoir modes to the system.
These coupling strengths are usually cast into the spectral density $J(\nu)= \half\sum_{\bf k}g_{\bf k}^2\delta(\nu -\Omega_{\bf k})$.
In the continuum limit the spectral density reads $J(\nu) \propto \nu^s$, where the exponent $s$ crucially determines the way a reservoir affects the system~\cite{Zhang2012, Breuer2016, Abdi2018}.
In this work we only study the ohmic bath ($s=1$) which is relevant in most physical systems.

We follow the standard procedure for deriving the master equation by assuming a separable total initial state $\chi(t_0) = \rho(t_0)\otimes R(t_0)$.
In the rotating wave approximation which is valid when $g_{\bf k} \ll \widetilde\omega, \Omega_{\bf k}$ the dynamics of the system in the case of single- and two-photon processes is described by the time-convolution-less master equation $\dot\rho = \supL_{nm}(t)\rho$.
The Liouvillian superoperator is given by
\begin{align}
\supL_{nm}(t)\rho \equiv
&-i[\omega a^\dag a, \rho]
+\tfrac{\Gamma_{nm}(t)}{2^n}\supD_{a^n}\rho
+\tfrac{\gamma_{nm}(t)}{2^n}\supD_{a^{\Dag n}}\rho \nonumber\\
&+\tfrac{\kappa_{nm}(t)}{2^n}(\supD'_{a^n}\rho +\supD'_{a^{\Dag n}}\rho), 
\label{master}%
\end{align}%
where $\supD_o\bullet \equiv 2o\bullet o^\dag -o^\dag o\bullet -\bullet o^\dag o$ is the standard Lindblad dissipator, while $\supD'_o\bullet \equiv 2o\bullet o -o o\bullet -\bullet o o$ is of a generalized one.
Here, we have introduced the renormalized frequency $\omega$, which indeed is the frequency that one observes in the lab.
The time-dependent damping rates $\gamma_{nm}(t),\Gamma_{nm}(t),\kappa_{nm}(t)$ vary with the bath correlation functions, see Appendix~\ref{app:rwamaster} for the detailed derivation and the explicit form of these parameters.
Because of the time dependence of the decay rates the above equation describes the non-Markovian evolution of the system.

\begin{figure}[tb]
\includegraphics[width=\columnwidth]{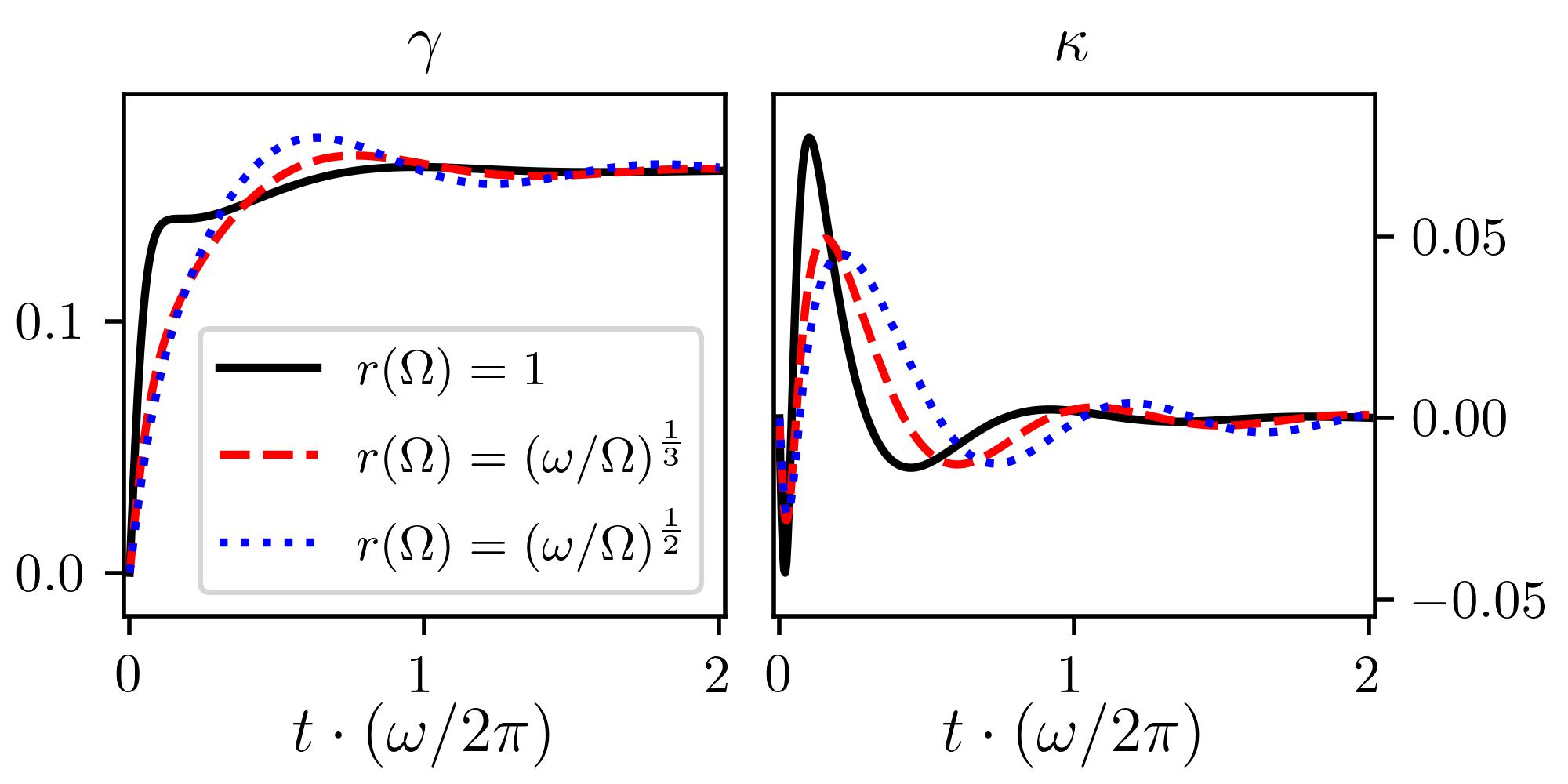}
\caption{The variations of bilinear ($n=m=1$) decay rates $\gamma$ (left) and $\kappa$ (right) with time when the environment is in a locally squeezed state with $r(\Omega)=1$ (black solid line), $r(\Omega) = (\omega/\Omega)^{-1/3}$ (red dashed line) and $r(\Omega) = (\omega/\Omega)^{-1/2}$ (blue dotted line).
Here, we have set $A=0.1$ and $\Omega_{\rm UV}=4\Omega_{\rm IR}=2\omega$, see the text for more information.}%
\label{fig:gamma11}%
\end{figure}

For computing these decay rates one needs to perform a frequency integration over the bath correlators and the spectral density function.
To avoid the divergences, one introduces ultraviolet (UV) and in some cases infrared (IR) cutoff frequencies.
Here, we shall need to consider both as it becomes clear shortly, and introduce the cutoffs in a soft way by multiplying the spectral density function with the exponential tails such that the spectral density read $J(\nu)= A\nu \exp\{-(\frac{\Omega_{\rm IR}}{\nu} +\frac{\nu}{\Omega_{\rm UV}})\}$, where $A$ is a dimensionless parameter that determines the system-reservoir interaction strength.
From a physical point of view the UV cutoff frequency is inversely related to the bath relaxation time, and thus, governs the bath memory effects.
In other words, when $\Omega_{\rm UV} \gg \omega$ the bath correlation functions have a characteristic time much shorter than that of the system.
Therefore, the reservoir approaches its steady-state too fast for the system to `feel' its dynamics.
In contrast, when $\Omega_{\rm UV}\sim\omega$ one expects to have visible effects resulting from the information back-flow.
In Fig.~\ref{fig:gamma11} the variations of the decay rates with time are shown for a squeezed reservoir when the bath operators in the interaction Hamiltonian is bilinear, i.e. $n=m=1$.

\subsection{Bath states}
As mentioned above and detailed in Appendix~\ref{app:rwamaster} the state of bath appears in the decay rates through its correlation functions.
These rates, in turn, determine the form and strength of the system-bath interactions.
Consequently, causing evolution of the system coherence.
In the number basis measurement the squeezing decay rate $\kappa$ is the only source of coherence build-up in the system through its interaction with the bath.
Indeed, the witness \eqref{witness} signals time-correlated occurrence of any coherence in the system which for the Fock basis measurement means any event that coherently and partially migrates the system from its initial state should result-in a nonzero $\calW_Q$, provided the measurements are invasive.
By looking at the explicit form of the dissipators $\supD$ and $\supD'$ one notices that unlike the standard Lindblad dissipators, the $\supD'$ ones may create superpositions in the number of bosons.
This may become clear by an example.
Assume that the system is initially in a given number state $\rho_{0}=\proj{n_0}$.
The first quantum stochastic modification imposed only by the dissipator $\supD_a$ is proportional to $d\rho_{\supD_{a}} \propto 2n_0 (\proj{n_0-2} -\proj{n_0})$, which is an incoherent redistribution of the system into the number states.
In contrast, for $\supD'_a$ contribution to the dynamics one finds $d\rho_{\supD'_a} \propto 2\sqrt{n_0(n_0+1)}\trans{n_0-1}{n_0+1} -\sqrt{n_0(n_0-1)}\trans{n_0-2}{n_0} -\sqrt{(n_0+1)(n_0+2)}\trans{n_0}{n_0+2}$.
This comprehensibly indicates accumulation of coherency in the system.

Hence, for a thermal bath where only the standard dissipators $\supD$ are effective the quantumness build up in the system when preparing and measuring it in the Fock basis is vanishing.
This statement generally holds for any bath with $\kappa(t)=0$.
For a squeezed bath $\kappa(t)\neq 0$ and depending on the microscopic properties of the bath modes one may even have a Markovian squeezed bath~\cite{Walls2008}.
However, this becomes the case only when the bath modes are interacting or in an entangled state.
For individually squeezed bath modes $\kappa(t\to\infty)\to 0$ because of the non-stationary nature of its bath correlator, and thus, the nonclassicality creation is a transient effect, which if long enough still measurable.
The focus of this work is put on such bath states.
Hence, only non-Markovian squeezed baths are studied.
Among which one may assume a uniformly squeezed bath where the modes are equally squeezed, i.e. the squeezing parameters is a constant $r(\Omega)=\text{const}$.
In general, $r(\Omega)$ is a function of mode frequency.
To take into account the effect of frequency dependence of the squeezing we consider $r(\Omega)\propto \Omega^\alpha$ with $\alpha=0,-\frac{1}{3},-\half$.
These choices of $\alpha$ allows us to easily regulate the frequency integrals in computing the decay rates and yet they are close to the relevant natural cases such as the relic gravitons~\cite{Grishchuk1990, Albrecht1994}, whose quantum effects has recently become of increasing interest, see e.g.~\cite{Kanno2021, Parikh2021}.

%
%
%----------BILINEAR----------%
\section{Bilinear interaction}\label{sec:bilinear}
We now present our numerical results for $\calW_Q$ computed in both number basis and a set of coherent states.
To better understand the effect of bath state and its memory we first investigate the simplest case of interaction were the system and reservoir modes interchange single particles.
Later and in the next section, we report the results for nonlinear interactions stemming from either the system or the reservoir operators.

\subsection{Parameters and numerical methods}\label{subsec:params}
For the number state basis we numerically solve for the master equation.
To do this we use the QuTiP package~\cite{Johansson2013}.
For having a rather fair comparison between different baths we choose the parameters such that their decay rates are comparable.
To this end, the parameters are adjusted for having a fixed value for the occupation number $\mean{b^{\Dag}(\Omega)b(\Omega)}$ at $\Omega=\omega$ in all cases.
By setting $r(\Omega)=1$ for the uniformly squeezed bath this corresponds to $r(\Omega)= (\Omega/\omega)^\alpha$ for the other two baths studied in this work.
To ensure a weak system-bath coupling we set $A=0.1$ in our numerical analyses.
Given the above mentioned bath parameters, our investigations show that with a Hilbert truncated at $d=20$ a relative error less that one percent is  guaranteed as long as initial states with $n_0\leq 5$ are employed in the numerics.
Finally, we set $\Omega_{\rm UV}=2\omega$ (and $\Omega_{\rm IR}=0.5\omega$) to make the memory effects appreciable and employ the time-dependent decay rates to solve the master equation \eqref{master}.

\subsection{Number state basis}
First we study the witness in the number state basis.
For this we consider different initial number states $\ket{n_0}$ with $n_0 \leq 5$ and various measurement times.
The optimal results for different system and bath states are found around equal time interval measurements.
Therefore, here we only report the results for $t_1=\tau$.
The overall behavior of $\calW_Q$ depends on the memory of the bath and its state.
The value of quantum measure is negligibly small and comparable to the numerical errors for the Markovian and thermal non-Markovian baths.
This results from the vanishing value of $\kappa$ in these cases as mentioned and discussed above.
That is, the quantumness build up from the standard dissipators $\supD$ as we have observed in the system interacting with a thermal or a Markovian bath is vanishing, when the system is prepared and measured in the Fock basis.
It is expected, however, that the quantum coherence be conveyed to the system from a squeezed bath since $\kappa(t)\neq 0$ for a finite time interval.
\begin{figure}[tb]
\includegraphics[width=0.49\columnwidth]{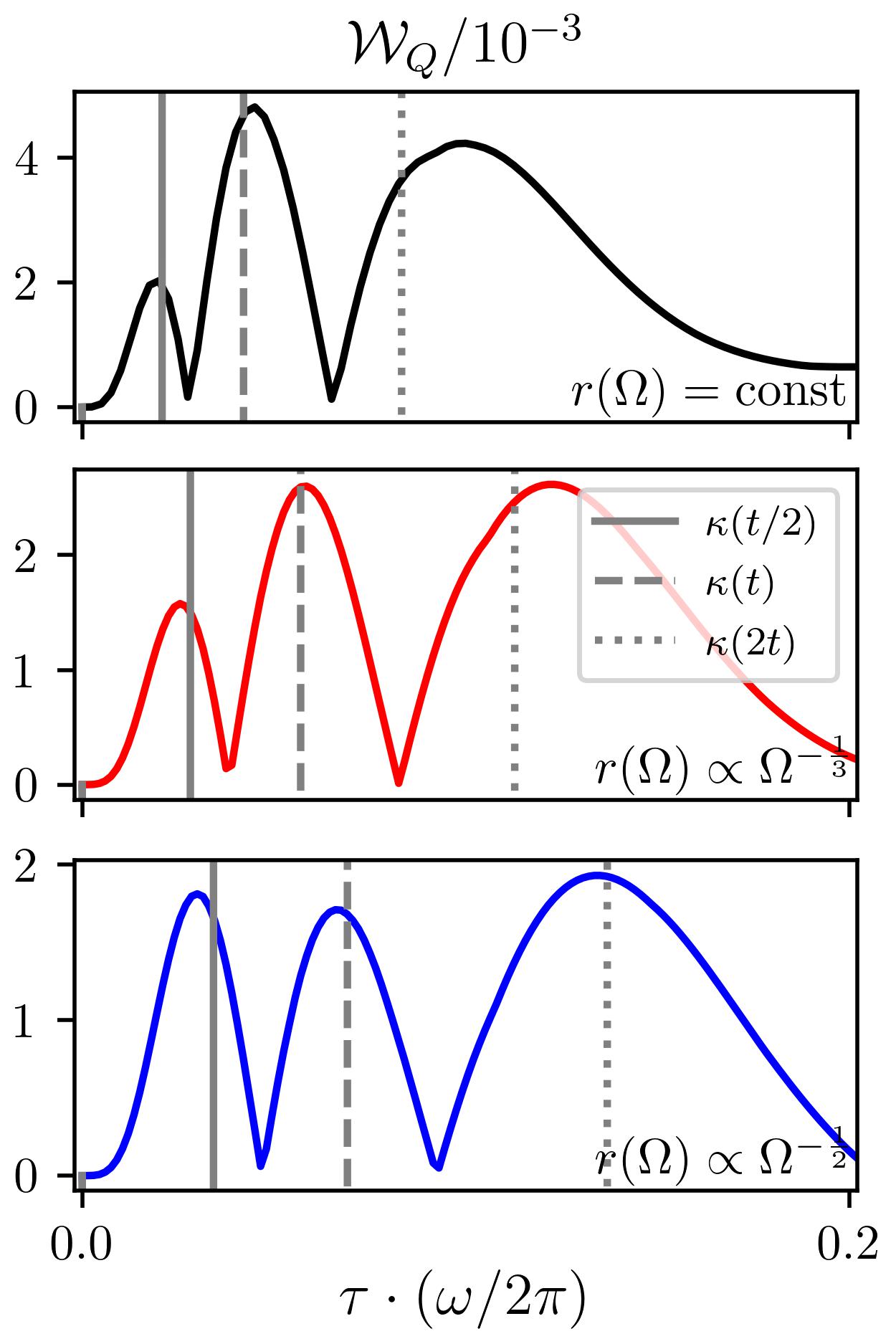}
\includegraphics[width=0.49\columnwidth]{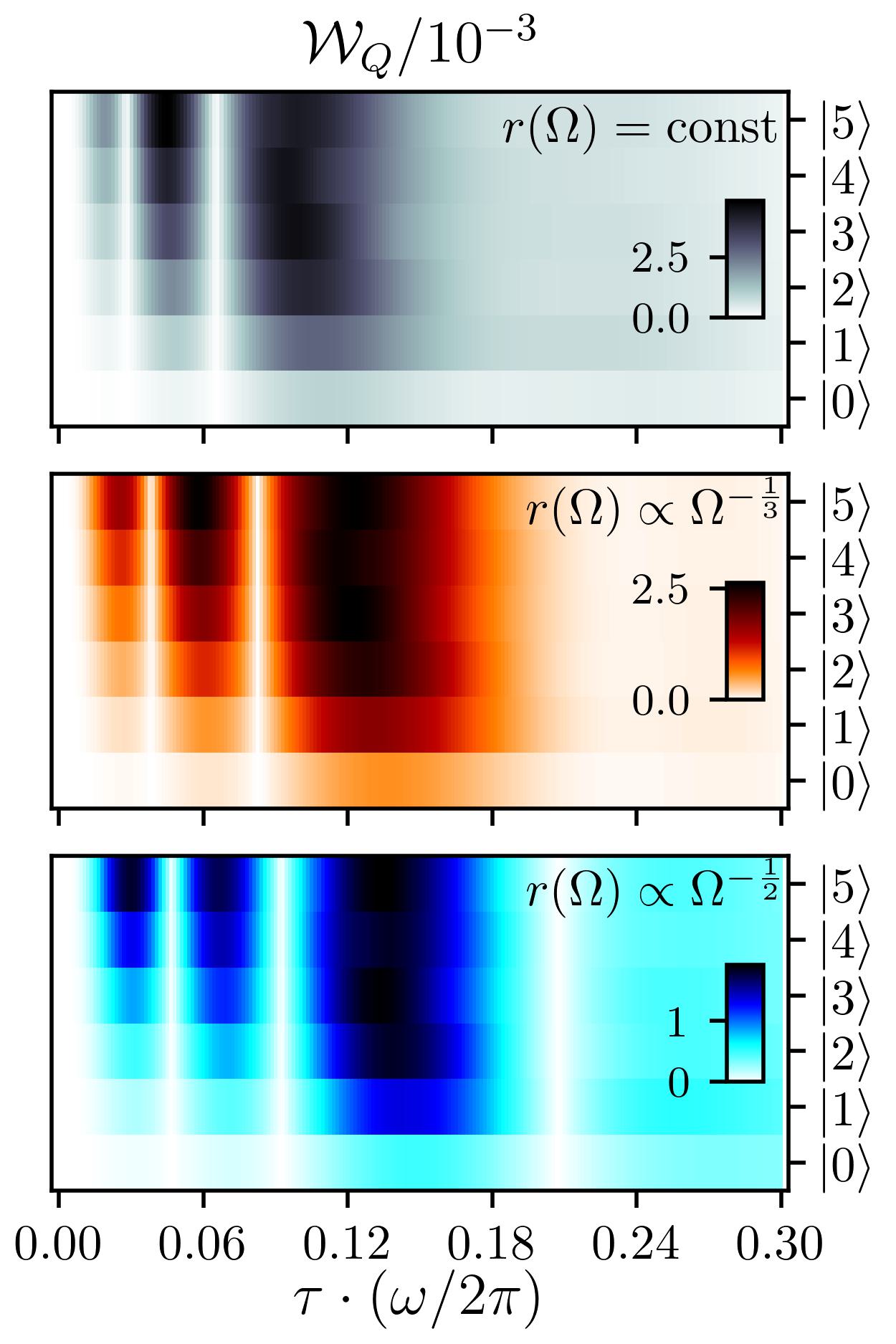}
\caption{Time variations of $\calW_Q$ for two evenly performed consecutive measurements ($t_1=\tau$) in the number state basis:
Left panels show the behavior for different squeezed baths when the system is initialized at the number state $\ket{5}$.
In right panels the variations are shown for different system initial states.
The gray lines indicate zeros of the damping rate $\kappa$ at three different time scales shown in the legend.
%The damping rates are the same as those in Fig.~\ref{fig:gamma11}.
}%
\label{fig:WQ11num}%
\end{figure}%

In Fig.~\ref{fig:WQ11num} the witness is plotted against $\tau=t_1$ for squeezed bath states with different properties discussed above, and for various initial system states.
Particularly, in the left panels the witness is plotted for the system initialized at $\ket{5}$.
The absolute value of the witness is small which stems from our weak coupling assumption, yet it retains some interesting features.
In the curves one notices several minima with vanishing $\calW_Q$ which obviously correspond to $P(n_2,t_2)=\sum_{n_1}P(n_2,t_2;n_1,t_1)$ for \textit{all} $n_2$ values.
The location of these minima and the following maxima depends on the time dependence of the squeezing decay rate $\kappa$, which gives an intuitive picture about the effect of non-Markovianity and the bath state properties.
Indeed, the maxima correspond to the critical points of $\kappa$, where the squeezed damping rate switches sign.
As a consequence, the minima stem from the change in the course of the coherence transfer at the times interrupted by the measurements $t=t_1,t_2$.
Hence, the probability $P(n_2,t_2)$ changes towards $\sum_{n_1}P(n_2,t_2;n_1,t_1)$ until they become equals.
In the figures we have indicated the relevant sign switch times with the gray lines.
Eventually, for long enough measurement times the measure approaches zero, which in turn stems from the overtaking of $\supD$ imposed dynamics, and thus, removing any coherence in the system.

The above discussion applies to all of the initial states studied in this work, i.e. for $\ket{n_0\leq 5}$.
Nonetheless, depending on the number of photons and the frequency dependence of the squeezing parameter each initial state has a different $\calW_Q$ profile.
In the right panels of Fig.~\ref{fig:WQ11num} we use density plots to present the value of $\calW_Q$ with $\tau=t_1$ for different initial states.
One clearly notices that the local minima with vanishing $\calW_Q$ occur exactly at the same time regardless of the initial state.
This further supports the effect of $\kappa$ sign switching and the measurement time interruptions.

\subsection{Phase space measurement}
\begin{figure}[tb]
\includegraphics[width=\columnwidth]{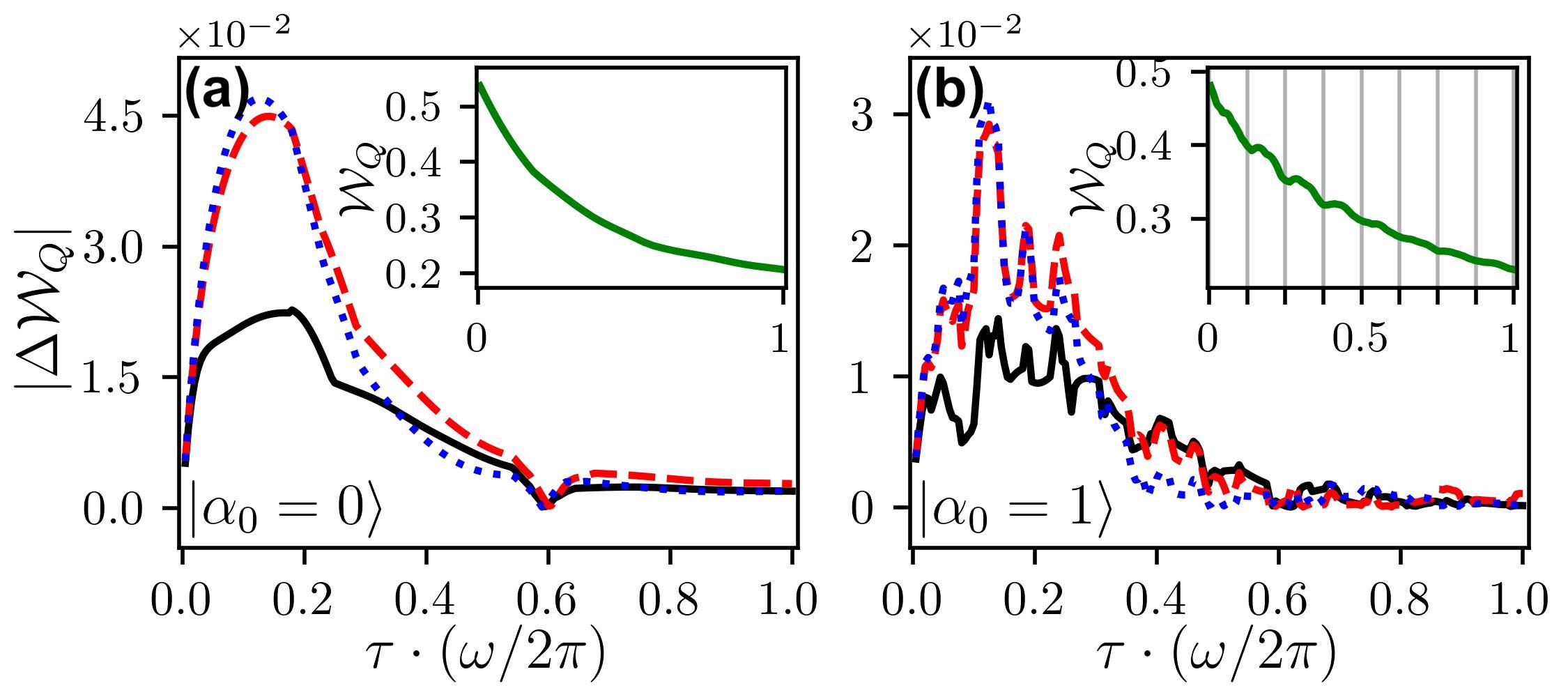}
\includegraphics[width=\columnwidth]{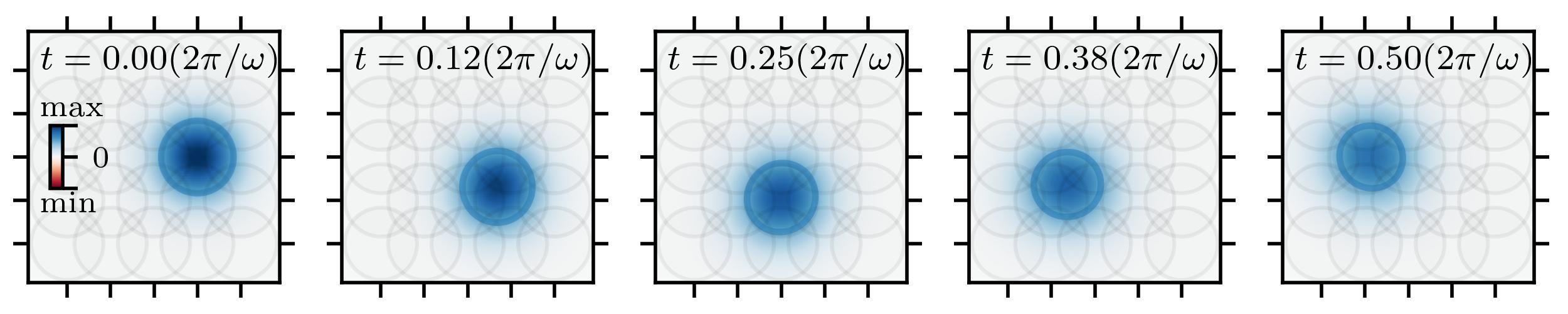}
\caption{Time variations of $|\Delta\calW_Q|$ for two evenly performed consecutive measurements ($t_1=\tau$) in the granulated coherent state basis with the initial state $\ket{\alpha_0=0}$ (a) and  $\ket{\alpha_0=1}$ (b) for squeezed baths with different squeezing parameters: $r(\Omega)=1$ (solid black line), $r(\Omega) =(\omega/\Omega)^{-1/3}$ (dashed red line), and $r(\Omega) =(\omega/\Omega)^{-1/2}$ (dotted blue line).
The $\calW_Q$ values for the Markovian bath are shown in the insets.
In the bottom panels the time evolution of the Husimi $Q$-function for the uniformly squeezed non-Markovian bath is shown for the initial state $\ket{\alpha_0=1}$ in the background of the chosen coherent basis.
The contours indicate the half of the maximum of the $Q$-function corresponding to the system state (blue) and the measurement basis (gray).
}%
\label{fig:WQ11coh}%
\end{figure}

Next we granulate the phase space and consider a set of coherent states distributed in the form of a square lattice as the basis for evaluating the quantum witness.
Given the computational resources at hand we only cover the phase space with 25 evenly spaced coherent states from $\ket{-2-2i}$ to $\ket{2+2i}$.
These basis states are shown by gray circles and shades in the lower panels of Fig.~\ref{fig:WQ11coh}.
The contours (circles) refer to the half maximum probability of the corresponding coherent state $Q$-function representation.
This choice of the spacing between the basis ensures that no poorly covered regions remain in the phase space and yet the overlaps remain reasonably small.
Moreover, we consider coherent initial states $\ket{\alpha_0}$ with $|\alpha_0|\leq 1$ because of the limited phase space coverage.
As one would expect the initial value of the measure of nonclassicality is nonzero, stemming from the probability overlap among the basis.
And its value decreases monotonically and approaches to a finite asymptotic value.
The non-vanishing asymptotic value is a consequence of limiting the phase space coverage to the the above mentioned range.
In other words, the information is gradually leaked into the phase space area which is not covered by the measurement basis, and thus, remains unaccessible.
This indeed is the general behavior observed for both Markovian and non-Markovian bath states.
However, there are slight differences given the bath state properties, which may prove useful for tracking the nonclassicality and non-Markovianity of a bath.
Therefore, we introduce the `deviation' quantity $\Delta\calW_Q \equiv \calW_Q^{\rm nM} -\calW_Q^{\rm M}$, where the superscripts `nM' and `M' indicate the non-Markovian and Markovian nature of the interacting bath, respectively.
%Indeed, one would get vanishing $\calW_Q$ values for a Markovian bath if there was no overlap among the computational basis and if they would cover the whole phase space.

The variations of $|\Delta\calW_Q|$ with $\tau=t_1$ are shown in Fig.~\ref{fig:WQ11coh}(a) and (b) for $\ket{\alpha_0=0}$ and $\ket{\alpha_0=1}$, respectively.
One notices that a non-Markovian bath can have a different effect on the system compared to the Markovian counterpart depending on its properties.
This is particularly more visible for the short time measurements and before the system approaching its asymptotic coherence.
In contrast to the number basis measurement scheme the squeezed bath with constant squeezing shows smaller deviation from the Markovian bath compared to the baths with frequency varying squeezing parameters.
In the case of $\ket{\alpha_0=1}$ we notice oscillations with $\tau$ (see the inset) which are believed to be artifact of the partial coverage of the phase space.
But the general behavior is similar to the vacuum state.
The oscillations can be better understood by tracking the time evolution of the coherent state in the phase space, which is visualized in the lower panels of Fig.~\ref{fig:WQ11coh}:
The minima happen as the system state approaches to the edges, either at $t=t_1$ or $t=t_2=2t_1$, that a smaller number of measurement basis states are present.

%
%
%----------NONLINEAR----------%
\section{Nonlinear interactions}\label{sec:nonlinear}
In this section we turn our study to the case of nonlinear interactions were the nonlinearity can have origins in either the system ($n=2$ but $m=1$) or the bath ($n=1$ but $m=2$).
%We skip the case of fully nonlinear interaction ($n=m=2$) as it does not offer results much different from the two former cases.
To distinguish these cases we adopt the notation $\calW_{n,m}$ for the quantum witness.
Similar to the bilinear case we consider both number basis and coherent state measurements as well as the baths considered above.
Moreover, to ensure a rather fair comparison between various cases the bath states are characterized by their squeezing parameter at the systems frequency, see Sec.~\ref{subsec:params}.
The corresponding decay rates are plotted in Fig.~\ref{fig:gammanm}.
One notices the effect of interaction type, whether it is single- or two-excitation exchange, and the squeezed bath properties, whether it is constant or a function of frequency, in the dissipation rates.

By contrasting the decay rates of nonlinear interactions with the linear case one my envision the general behavior of the quantum witness $\calW_{n,m}$ from the previously studied bilinear case.
Even though this could give a generally acceptable deduction, but the microscopic form of the interaction also plays an important role in the measurable quantumness.
The two-excitation absorption/emission process on the bath side is only reflected in the decay rate values, while the dissipators retain their form of the bilinear interaction.
Nonetheless, when nonlinearity of the interaction roots back to the system and double excitations in system are created or annihilated the dissipators become nonlinear.
This nonlinearity reflects in the behavior of the witness and its dramatic dependence on the bath properties and the system initial state, see below.

\subsection{Number state basis}
\begin{figure}[tb]
\includegraphics[width=0.495\columnwidth]{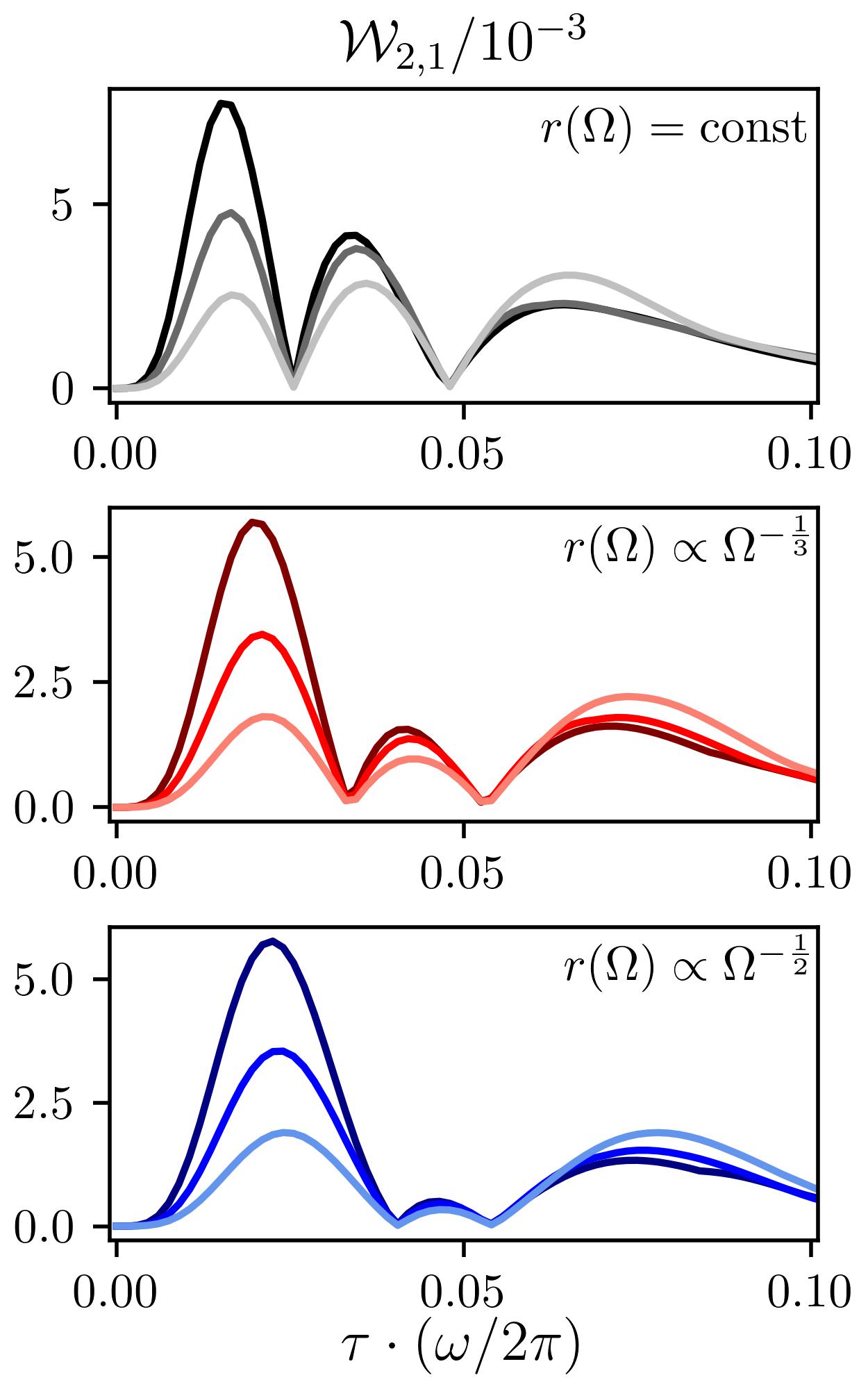}
\includegraphics[width=0.495\columnwidth]{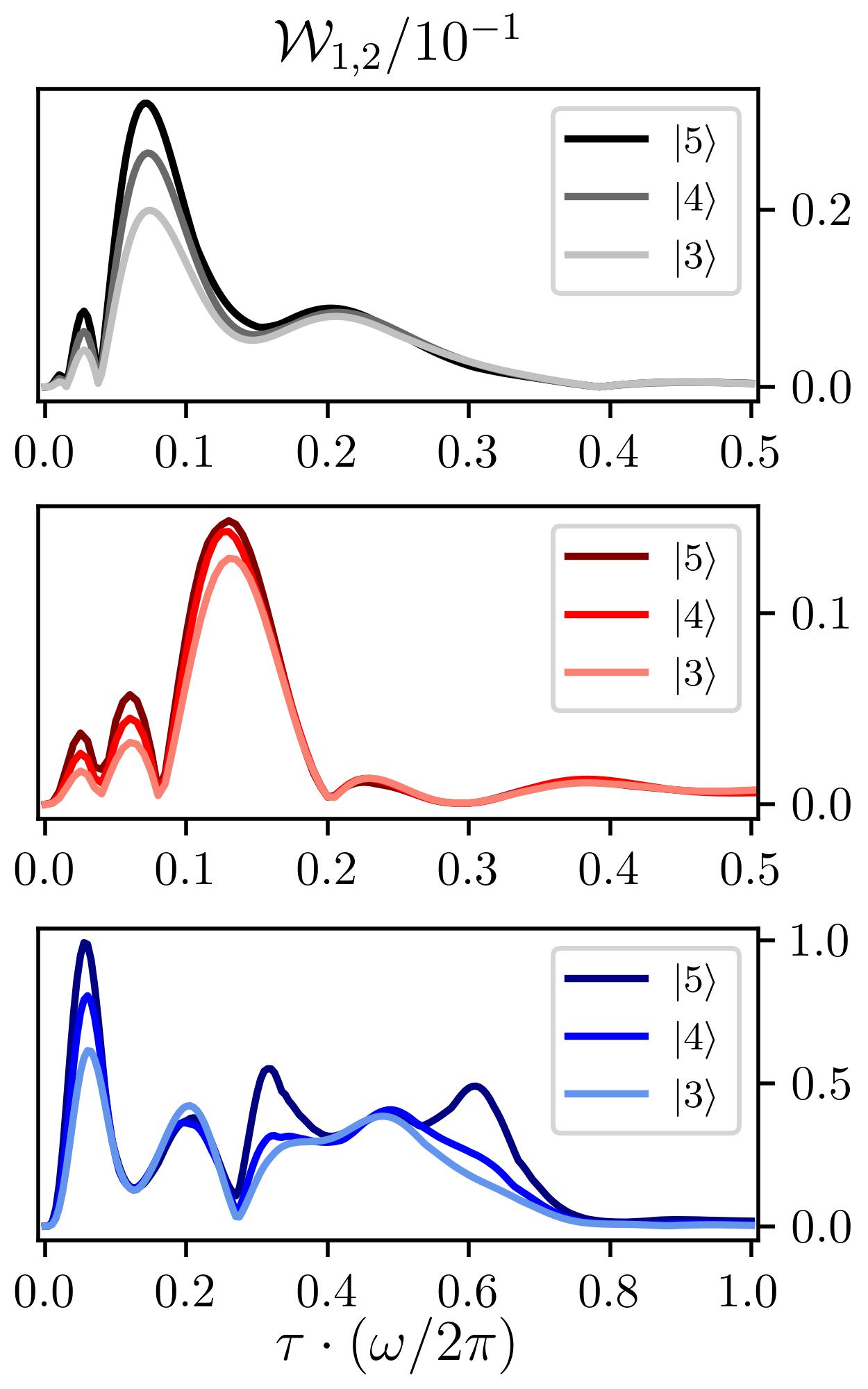}
\caption{The quantumness of a system with different initial states in nonlinear interaction with a squeezed bath:
The left (right) columns present the value of $\calW_{2,1}$ ($\calW_{1,2}$).
In (a), (b) the squeezing parameter is constant $r(\Omega)=1$, in (c), (d) it is $r(\Omega)=(\omega/\Omega)^{\frac{1}{3}}$, and in (e), (f) one had $r(\Omega)=(\omega/\Omega)^{\frac{1}{2}}$.
We have set $\tau = t_1$ and the decay rates drawn in Fig.~\ref{fig:gammanm} are used in the numerics.
}%
\label{fig:WQnmnum}%
\end{figure}

We first turn to the case were two excitations of system are exchanged with a single excitation in the bath.
Again, we consider a squeezed bath with different frequency dependence of their squeezing parameters $r(\Omega)$.
However, note that unlike the bilinear case the asymptotic decay rate values of these baths are not the same anymore [see Appendix~\ref{app:rwamaster}].
Despite the nonlinear nature of the $\supD$ dissipators in this case our numerical results show that no appreciable quantumness is conveyed to a system initialized in the number basis.
And similar to the bilinear case it is the squeezing decay rates ($\kappa$) responsibility to create a coherence in the system.
Even though the way they affect the system is different.
We observe the same correspondence between the local maxima of $\calW_{2,1}$ and the times at which $\kappa(t)$ changes sign.
But the height of the peaks at different system initial states behaves differently when compared to the bilinear case.
The results are summarized in Fig.~\ref{fig:WQnmnum}.

In the bilinear case one observes a monotonic increase in the witness $\calW_{1,1}=\calW_Q$ as the number of excitation in the initial state increases [see the right panels in Fig.~\ref{fig:WQ11num}].
This behavior holds for all three local maxima occurring at the measurement time period corresponding to $\tau \in [0,0.2(2\pi/\omega)]$ and $t_1=\tau$.
In contrast, such a monotonic manner does not hold for the nonlinear dissipators that emerge in the $n=2,m=1$ case.
This is such that while $\calW_{2,1}$ is a rapidly increasing function of $n_0$ at the shorter time measurements, it becomes a decreasing function when the outcomes of the measurements at longer times are inserted into the calculations.
The explicit form of the dissipators give an intuitive picture about the nature of this behavior.
Indeed the dissipators have larger effects on the system when its excitation number is larger.
That is, in the language of quantum trajectories method for the system initialized at $\rho_0=\proj{n_0}$ the quantum jump due to the dissipators is $|d\rho|\propto n_0^2$.
Therefore, in the longer time evolution and measurements one expects larger effect on the system with higher initial excitation numbers.
Nevertheless, this holds for both $\supD$ that tend to decohere the system and $\supD'$ that create coherence.
On the other hand, the larger values for $\gamma$ and $\Gamma$ overtake the constructive effect of $\kappa$, and thus, the witness trend reduces faster for higher initial excitations.

When the reservoir modes exchange two excitations with one from the system the created quantumness experiences a dramatic enhancement compared to the bilinear ($n=m=1$) and the opposite ($n=2,m=1$) cases.
The smaller value of the decoherence rates $\gamma$ and $\Gamma$, while persistent and larger values for $\kappa$ explain this larger value as well as the general behavior of $\calW_{1,2}$ with the measurement time $\tau$.
The witness for the bath with squeezing parameter $r(\Omega)=(\omega/\Omega)^{1/2}$ retains an appreciable value even for measurement times as long as $t_1=\tau=\pi/\omega$ despite rather large values of the decoherence rates.

\subsection{Phase space measurement}
For the measurements performed in the coherent state basis at the granulated phase space, we again consider the deviations of witness from the Markovian bath.
Note that in the case of nonlinear system-reservoir interaction the Markovian bath varies for the baths with different state properties which here is the squeezing parameters dependence on frequency.
Consequently, when computing $\Delta\calW$ for each case we take into account their corresponding Markovian reservoirs.

\begin{figure}[tb]
\includegraphics[width=\columnwidth]{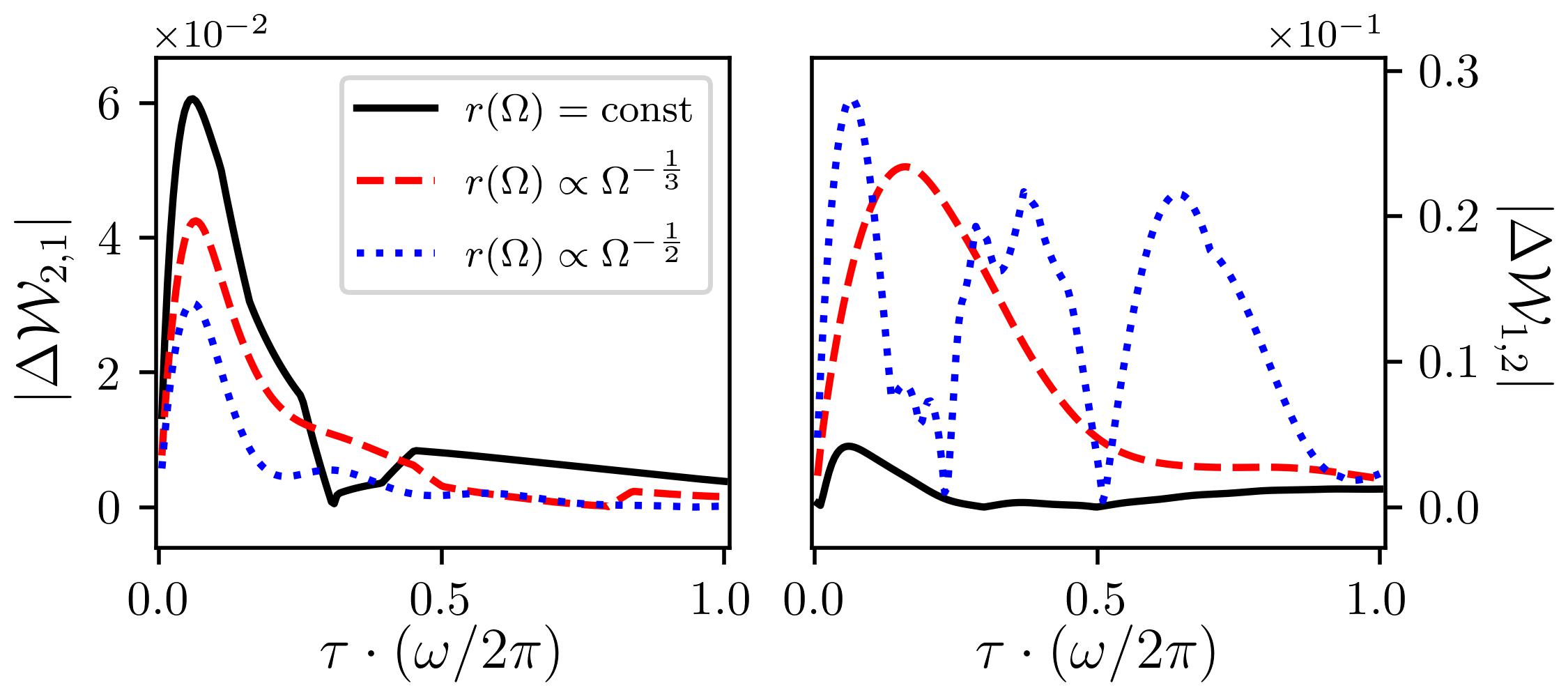}
\includegraphics[width=\columnwidth]{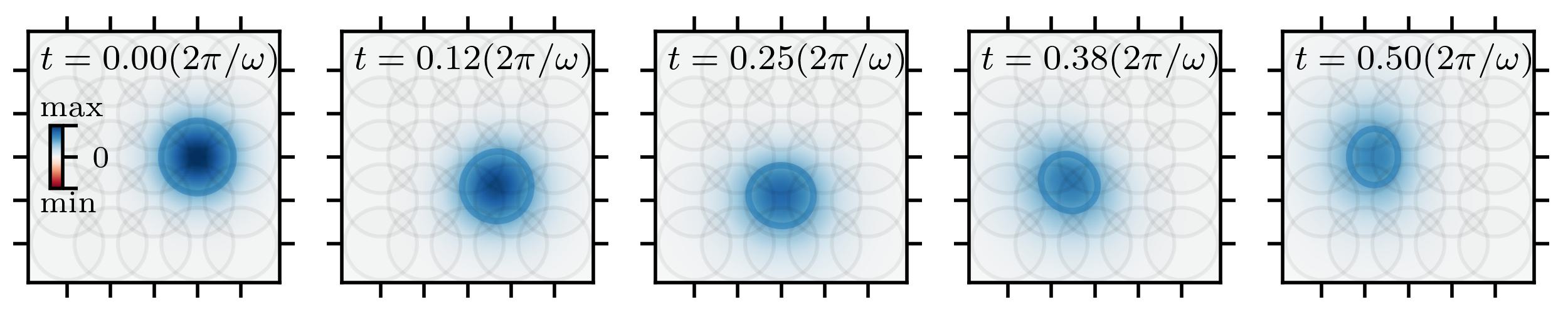}
\includegraphics[width=\columnwidth]{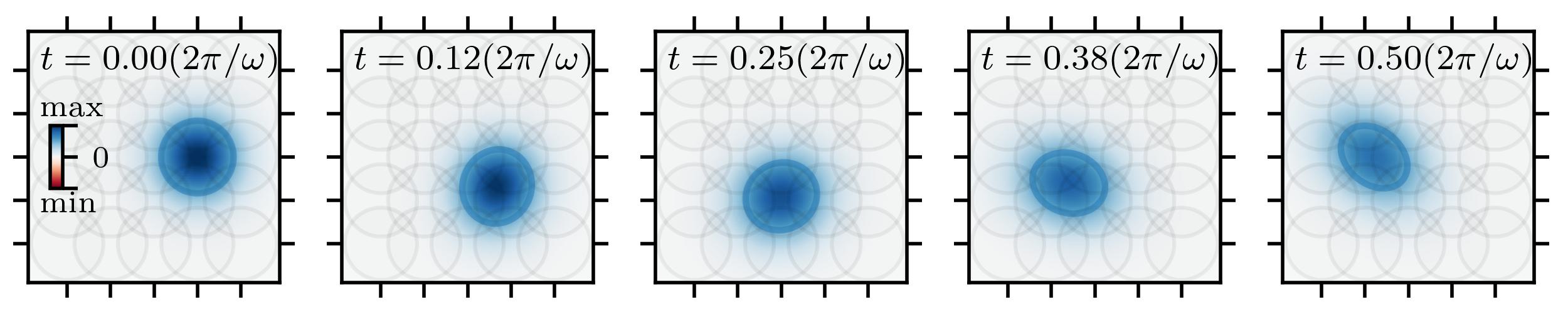}
\caption{Time variations of $|\Delta\calW_{n,m}|$ for two evenly performed consecutive measurements ($t_1=\tau$) in the granulated coherent state basis with the initial state $\ket{\alpha_0=0}$ for squeezed baths with different squeezing parameters (see the legend).
In the bottom panels the time evolution of the Husimi $Q$-function for the uniformly squeezed non-Markovian bath is shown for the initial state $\ket{\alpha_0=1}$ in the background of the chosen coherent basis.
}%
\label{fig:WQnmcoh}%
\end{figure}

In Fig.~\ref{fig:WQnmcoh} we plot the variations of the `pure witness' $|\Delta\calW_{n,m}|$ with respect to the measurement time when the system is initially in the vacuum state.
Interestingly, these two nonlinear interaction cases impart quantum effects in a way similar to those observed in the number basis measurement.
This includes short-time and comparable effects for $\Delta\calW_{2,1}$ as well as long-time oscillations of $\Delta\calW_{1,2}$ when $r(\Omega)=(\omega/\Omega)^{1/2}$.
But one also notices differences between the number and coherent basis measurement schemes such as the amplitude of the measure [compare Fig.~\ref{fig:WQnmcoh}(b) with the right panels in Fig.~\ref{fig:WQnmnum}].
Such similarities and differences resulting from the measurement basis can be exploited for better identifying the nature of a reservoir and its interaction with the probe.

%
%
%----------CONCLUSION----------%
\section{Summary and outlook}\label{sec:conclusion}
We have studied the possibility of tracking the quantum features impinged into a system through its interaction with a squeezed reservoir with non-negligible memory effects.
We have considered three different interaction schemes between the system and bath modes.
That is, the bilinear and two nonlinear excitation exchange schemes between the system and bath modes.
We also take into account the possibility of having squeezed baths with squeezing parameters that is a function of frequency.
We employ the well-defined coherence witness for investigating the probability of a system in this scheme behaving quantum mechanically.
For doing so we consider two measurement bases, the Fock basis and phase space measurement.
To avoid divergencies in the latter which stems from the over-completeness of the coherent state, we propose to granulated the interested phase space area.
Our findings suggest that either of the interaction schemes (bilinear or nonlinear) as well as the bath properties (in this work the squeezing parameter $r(\Omega)$) has a different fingerprint.
Those features can be partially revealed by performing measurements in the number basis or in the phase space.
And that in the case of number basis measurements the initial state of the system plays an important role in revealing the nature of the interaction, whether it is bilinear, nonlinear with double-excitation exchanges at the system side, or nonlinear with the double-excitation exchanges in the reservoir modes.

In this work we have only studied the case of an environment with Ohmic spectral density, which is the most convenient one in the theory of open quantum systems.
One could go beyond this particular case and investigate the fingerprints of various spectral densities in the measure of nonclassicality.
Moreover, one could also envisage having system-bath interactions in the form of scattering.
That is, interactions where in the Hamiltonian Eq.~\eqref{hamiltonian} one has $q \propto (a^{\Dag}a)^n$ and/or $Q_{\bf k} \propto (b_{\bf k}^{\Dag}b_{\bf k}^{})^m$.
In the former case where the system quanta are scattered, one expects to have a nonclassicality built-up in the number basis measurement.
This can be seen from the number preserving form of the system operator in the interaction Hamiltonian.
In other words, one has $[a^{\Dag}a,q \sum_{\mathbf{k}} g_{\bf k} Q_{\bf k}]=0$ if $q \propto (a^{\Dag}a)^n$.
Therefore, the resulting dissipator does not redistribute the system quanta and thus a vanishing $\calW_Q$.
When the scattering occurs for the reservoir modes and $q\propto a^n +a^{\Dag n}$ we expect to have nonzero values for the measure for all bath states.
Nevertheless, a detailed understanding of these cases invokes a dedicated study.

%
%
%----------ACKNOWLEDGEMENTS----------%
%\begin{acknowledgements}
%We are just very grateful.
%\end{acknowledgements}

\onecolumngrid
\appendix
%\begin{widetext}

%%%%%%%%%%%%%%%%%%%%%%%%%%%%%%%%%%%%%%%%%%%%%%
\setcounter{equation}{0}
\setcounter{figure}{0}
\setcounter{table}{0}
\makeatletter
\renewcommand{\theequation}{A\arabic{equation}}
\renewcommand{\thefigure}{A\arabic{figure}}
\section{Time-convolutionless non-Markovian master equation}\label{app:master}
A general non-Markovian master equation is derived by the method of projection operators.
In our following analysis we assume that:
i) The initial system-reservoir state is separable $\chi(t_0)=\rho(t_0)\otimes R(t_0)$.
ii) The state of reservoir is not appreciably affected $R(t)\approx R(t_0)$.
iii) The state of reservoir is zero-mean Gaussian.
The projection to the `relevant' part of the density matrix, the system, is provided by the operator $\supP\chi = \tr{\rm B}\{\chi\}\otimes R \equiv \rho$.
Its complementary is thus $\supQ=\one -\supP$, where $\one$ is the identity operator.
Our aim is to find a time-convolutionless master equation for the relevant part of the master equation in the form of
\begin{equation}
\frac{d}{dt}\supP\chi(t) = \calK(t)\supP\chi(t).
\label{tclmaster}
\end{equation}
In the interaction picture the dynamics is given by
\begin{equation}
\dot\chi = -i\big[H_{\rm int}(t),\chi \big]=\supL(t)\chi,
\end{equation}
where $H_{\rm int}(t)=\exp\{i H_0 t\}H_{\rm int}\exp\{-i H_0 t\}$ with $H_0=\omega a^\dag a^{} +\sum_{\bf k} \Omega_{\bf k}b_{\bf k}^\dag b_{\bf k}^{}$ and $H_{\rm int}=q \sum_{\mathbf{k}} g_{\bf k} Q_{\bf k}$ is the Hamiltonian in the interaction picture and we have introduced the Liouvillian superoperator $\supL$.
The Kernel is found as~\cite{Breuer2007}
\begin{equation}
\calK(t)=\supP\supL(t)\Big\{\one -\int_0^t \!ds\big[\supT_-e^{\int_s^t d\tau \supQ\supL(\tau)}\big] \supQ\supL(s)\supP \big[\supT_+e^{-\int_s^t d\tau \supL(\tau)}\big] \Big\}^{-1}\supP,
\end{equation}
where $\supT_+$($\supT_-$) is the time ordering (anti-time ordering) operator.
Under the given assumptions Eq.~\eqref{tclmaster} is an exact equation which gives the non-Markovian evolution of the system density matrix.
Nonetheless, its complexity is not less than the von Neumann equation itself.
To make it tractable one, thus, turns into a perturbative treatment, where the kernel $\calK(t)$ is perturbatively expanded in terms of different orders of $\supL(t)$ based on the weak coupling of the system and reservoir.
In terms of the spectral density function this holds for $A \ll 1$.
We note that $(1-x)^{-1}=\sum_{n=0}^\infty x^n$ and $\supP^2=\supP$.
Hence
\begin{equation}
\calK(t)=\sum_{n=0}^\infty \supP\supL(t)\Big\{\int_0^t \!ds\big[\supT_-e^{\int_s^t d\tau \supQ\supL(\tau)}\big] \supQ\supL(s)\supP \big[\supT_+e^{-\int_s^t d\tau \supL(\tau)}\big] \Big\}^n.
\end{equation}
Within this one has
\begin{align}
\supT_+\exp\{\int_s^t d\tau\supL(\tau)\} &= \sum_{k=0}^\infty \int_s^t\hspace{-2mm}d\tau_1\int_{\tau_1}^t\hspace{-1mm}d\tau_2\cdots\int_{\tau_{k-1}}^t\hspace{-3mm}d\tau_k\ \supL(\tau_1)\supL(\tau_2)\cdots \supL(\tau_k), \nonumber\\
\supT_-\exp\{\int_s^t d\tau\supL(\tau)\} &= \sum_{k=0}^\infty \int_s^t\hspace{-2mm}d\tau_1\int_s^{\tau_1}\hspace{-3mm}d\tau_2\cdots\int_s^{\tau_{k-1}}\hspace{-3mm}d\tau_k\ \supL(\tau_1)\supL(\tau_2)\cdots \supL(\tau_k).
\end{align}
By exploiting the fact that the reservoir is in a zero-mean Gaussian state one has $\supP\supL(t)\supP=0$ and for all odd multi-time moments.
Moreover, any even multi-time moment can be expressed as a sum over products of second moments.
Therefore, to the second order the kernel reads
\begin{equation}
\calK_2(t) = \int_0^t\!ds\ \supP\supL(t)\supL(s)\supP.
\end{equation}
We now use the explicit form of the Liouvillian to derive the master equation.
For this, we recall that
\begin{equation}
H_{\rm int} =\sum_{\bf k}g_{\bf k}q Q_{\bf k}
\end{equation}
%where we have introduced the reservoir operator $Q \equiv \sum_{\bf k}g_{\bf k}(b_{\bf k}^{} +b_{\bf k}^\dag)$.
The interaction picture Liouvillian then reads
\begin{equation}
\supL(t)\chi = -i\Big[\sum_{\bf k}g_{\bf k}q(t) Q_{\bf k}(t),\chi \Big]
\end{equation}
Therefore, for the second order kernel we find
\begin{align}
\calK_2(t)\chi(t) &= -\int_{t_0}^t\!ds\ \Big\{ \calC^\Re(t,s)\big[q(t),[q(s),\rho(t)]\big]
+i\ \calC^\Im (t,s)\big[q(t),\{q(s), \rho(t)\}\big] \Big\}\otimes R(t_0),
\label{kernel2}
\end{align}
where we have introduced the bath correlation function
\begin{equation}
\calC(t,t') = \sum_{\bf k}\sum_{\mathbf{k}'}g_{\bf k}g_{\mathbf{k}'}\tr{\rm B}\{R(t_0) Q_{\bf k}(t)Q_{\mathbf{k}'}(t')\}.
\end{equation}
Since the bath operators in the interaction are taken as $Q_{\bf k} = (b_{\bf k}^m +b_{\bf k}^{\dag m})/\sqrt{2^m}$ with $m=1,2$ we arrive at
\begin{equation}
\calC(t,t') \equiv \calC_m(t,t') = \frac{1}{2^m}\sum_{\bf k}\sum_{\mathbf{k}'}g_{\bf k}g_{\mathbf{k}'}
\tr{\rm B}\Big\{
R(t_0)
\big(b_{\bf k}^me^{-im\Omega_{\bf k}t} +b_{\bf k}^{\dag m} e^{im\Omega_{\bf k}t}\big) 
\big(b_{\mathbf{k}'}^m e^{-im\Omega_{\mathbf{k}'}t'} +b_{\mathbf{k}'}^{\dag m} e^{im\Omega_{\mathbf{k}'}t'}\big)
\Big\},
\end{equation}
with $\Re$ and $\Im$ representing its real and imaginary parts, respectively.

\subsection{Thermal bath}
\subsubsection{$m=1$}
For a thermal bath, when the bath operators are linear one finds
$$\calC_1(t,t') = \half\sum_{\bf k}g_{\bf k}^2
\Big\{
\Nbos(\Omega_{\bf k})e^{i\Omega_{\bf k}(t-t')} +\big[\Nbos(\Omega_{\bf k})+1 \big]e^{-i\Omega_{\bf k}(t-t')}
\Big\}.$$
Since we have $\Nbos(\omega)=(e^{\beta\hbar\omega}-1)^{-1}$ the above expression is rewritten as
\begin{equation}
\calC_1(t,t') = \half\sum_{\bf k}g_{\bf k}^2
\Big\{
\coth\big(\half\beta\hbar\Omega_{\bf k}\big)\cos\big[\Omega_{\bf k}(t-t')\big] -i\sin\big[\Omega_{\bf k}(t-t')\big]
\Big\}.
\end{equation}
In the continuous limit one does the substitution $\half\sum_{\bf k}g_{\bf k}^2 \to \int d\Omega J(\Omega)$, where we have introduced the spectral density $J(\Omega)$ and dropped the subscript $\mathbf{k}$ from the frequencies.
The thermal bath correlator thus reads
\begin{equation}
\calC_1(t,t') = \int_0^\infty\! d\Omega\ J(\Omega)
\Big\{
\coth\big(\half\beta\hbar\Omega\big)\cos\big[\Omega(t-t')\big] -i\sin\big[\Omega(t-t')\big]
\Big\},
\end{equation}
To avoid the ultraviolet catastrophe one needs to regulate the spectral density.
Here, we employ a soft exponential tail and set the spectral density to $J(\nu)=A\Omega_{\rm c}^{1-s}\nu^s e^{-\nu/\Omega_{\rm c}}$ with the cutoff frequency $\Omega_{\rm c}$.

\subsubsection{$m=2$}
In the case of two-photon excitations in the bath one instead arrives at
$$\calC_2(t,t') = \tfrac{1}{4}\sum_{\bf k}g_{\bf k}^2
\Big\{
\Nbos(\Omega_{\bf k})\big[\Nbos(\Omega_{\bf k}) -1 \big] e^{2i\Omega_{\bf k}(t-t')} +\Big[\Nbos(\Omega_{\bf k})\big[\Nbos(\Omega_{\bf k})+3\big]+2 \Big]e^{-2i\Omega_{\bf k}(t-t')}
\Big\}.$$
After a rearrangement and for a continuum of bath modes we have
\begin{equation}
\calC_2(t,t') = \int_0^\infty\! d\Omega\ J(\Omega)
\Big\{
\tfrac{1}{4}\csch^2(\half\beta\hbar\Omega)\big(2\cosh(\beta\hbar\Omega) -1\big)\cos\big[2\Omega(t-t')\big] -i\coth\big(\half\beta\hbar\Omega\big)\sin\big[2\Omega(t-t')\big]
\Big\}.
\end{equation}

\subsection{Squeezed vacuum bath}
\subsubsection{$m=1$}
We now consider the reservoir in a \textit{locally} squeezed vacuum state $R(t_0)=\proj{\xi}$ with $\ket\xi = \bigotimes_{\bf k}S_{\bf k}(\xi_{\bf k})\ket{0}_{\bf k}$ where the squeezing operator of each mode is $S_{\bf k}(\xi_{\bf k})=\exp\{\half(\xi_{\bf k}^*b_{\bf k}^2 -\xi_{\bf k}b_{\bf k}^{\dag 2})\}$ with the squeezing amplitude $r_{\bf k}$ and angle $\theta_{\bf k}$ encoded in $\xi_{\bf k}=r_{\bf k}e^{i\theta_{\bf k}}$.
Hence, one arrives at
\begin{equation}
\calC_1(t,t') = \half\sum_{\bf k} g_{\bf k}^2
\Big\{
\half\sinh 2r_{\bf k} \big[e^{-i\Omega_{\bf k}(t+t')}e^{i\theta_{\bf k}} +e^{i\Omega_{\bf k}(t+t')}e^{-i\theta_{\bf k}}\big]
+(1 +\sinh^2\! r_{\bf k})e^{-i\Omega_{\bf k}(t-t')}
+\sinh^2\! r_{\bf k} e^{i\Omega_{\bf k}(t-t')}
\Big\}.
\end{equation}
This can be reorganized into the following form
\begin{equation}
\calC_1(t,t') = \half\sum_{\bf k} g_{\bf k}^2
\Big\{
\cos[\Omega_{\bf k}(t-t')]\cosh[2r(\Omega_{\bf k})]
+\cos[\Omega_{\bf k}(t+t')-\theta(\Omega_{\bf k})]\sinh[2r(\Omega_{\bf k})]
-i\sin[\Omega_{\bf k}(t-t')]
\Big\}.
\end{equation}
In the continuum limit this reads
\begin{equation}
\calC_1(t,t') = \int_0^\infty\! d\Omega\ J(\Omega)
\Big\{
\cos[\Omega(t-t')]\cosh[2r(\Omega)]
-\cos[\Omega(t+t')-\theta(\Omega)]\sinh[2r(\Omega)]
-i\sin[\Omega(t-t')]
\Big\}.
\end{equation}
Note that, by assuming a local squeezing we rule out the existence of two-mode squeezed states that could give rise to stationary squeezing noise, see e.g. Ref.~\cite{Walls2008}.

\subsubsection{$m=2$}
The two-photon counterpart gives
\begin{align}
\calC_2(t,t') = \int_0^\infty\! d\Omega\ J(\Omega)
&\Big\{
\tfrac{1}{8}\cos[2\Omega(t-t')](7 +\cosh[4r(\Omega)]) \nonumber\\
&-\tfrac{3}{4}\cos[2\Omega(t+t')-2\theta(\Omega)]\sinh^2[2r(\Omega)]
-i\sin[2\Omega(t-t')]\cosh[2r(\Omega)]
\Big\}.
\end{align}

\subsection{The master equation}
Having the bath correlators settled, we now turn to the master equation.
%One may choose using the canonical formalism ($[q,p]=i$) or the quantum optical convention ($[q,p]=\half i$). It seems the canonical thing can give a fairer comparison for the power of the two baths.
We choose the system operator such that $q=\sqrt{2^{-n}}(a^n +a^{\dag n})\equiv q_n$ with $n=1,2$.
This allows for a fair comparison of the two kind of baths.
Therefore, one has $q_n(t)=\sqrt{2^{-n}}\big(a^n e^{-in\omega t} +a^{\dag n} e^{in\omega t} \big) = \sqrt{2^{-n}}(a^n +a^{\dag n})\cos n\omega t -\sqrt{2^{-n}}i(a^n -a^{\dag n})\sin n\omega t \equiv q_n\cos n\omega t +p_n\sin n\omega t$, where we have introduced $p_n \equiv -\sqrt{2^{-n}}i(a^n -a^{\dag n})$.
After some tedious calculations one arrives at the following master equation in the Schrodinger picture
\begin{align}
\dot\rho (t) =
&-i[H_{\rm s}, \rho]
-i\widetilde\Lambda_{n,m}(t) [q_n^2, \rho]
-i\widetilde\Gamma_{n,m}(t) \big[q_n,\{p_n,\rho\}\big]
-\widetilde\gamma_{n,m}(t) \big[q_n,[q_n,\rho]\big]
+\widetilde\lambda_{n,m}(t) \big[q_n,\{p_n,\rho\}\big],
\label{masterfull}
\end{align}
where we have introduced the following ($t_0=0$)
\begin{subequations}
\begin{align}
\widetilde\Lambda_{n,m}(t) &\equiv \int_{0}^t\!ds\ \calC_m^\Im(t,s)\cos[n\omega (t-s)], \\
\widetilde\Gamma_{n,m}(t) &\equiv \int_{0}^t\!ds\ \calC_m^\Im(t,s)\sin[n\omega (t-s)], \\
\widetilde\gamma_{n,m}(t) &\equiv \int_{0}^t\!ds\ \calC_m^\Re(t,s)\cos[n\omega (t-s)], \\
\widetilde\lambda_{n,m}(t) &\equiv \int_{0}^t\!ds\ \calC_m^\Re(t,s)\sin[n\omega (t-s)].
\end{align}%
\label{params}%
\end{subequations}%
%In Fig.~\ref{fig:params} the evolution of these parameters is plotted for three different states.
%
%\begin{figure}[tb]
%\includegraphics[width=0.5\columnwidth]{Figs/vac1.jpg}
%\includegraphics[width=0.5\columnwidth]{Figs/thr1.jpg}
%\includegraphics[width=0.5\columnwidth]{Figs/sqz1.jpg}
%\caption{Time evolution of the parameters in the non-Markovian master equation for: Vacuum state (top), thermal state with $\beta=0.1$ (middle), and squeezed state with $r=1$ (bottom).
%Here, the cutoff frequency is set to $\Omega_{\rm c}=10\omega$.}
%\label{fig:params}%
%\end{figure}
%

%%%%%%%%%%%%%%%%%%%%%%%%%%%%%%%%%%%%%%%%%%%%%%
\setcounter{equation}{0}
\setcounter{figure}{0}
\setcounter{table}{0}
\makeatletter
\renewcommand{\theequation}{B\arabic{equation}}
\renewcommand{\thefigure}{B\arabic{figure}}
\section{Rotating wave approximation}\label{app:rwamaster}
In the weak interaction regime that we are interested in, one applies the rotating wave approximation and arrives at a master equation in terms of the creation and annihilation operators.
Hence, we now simplify the second order kernel \eqref{kernel2} by applying the rotating wave approximation.
\begin{figure}[tb]
\includegraphics[width=0.5\columnwidth]{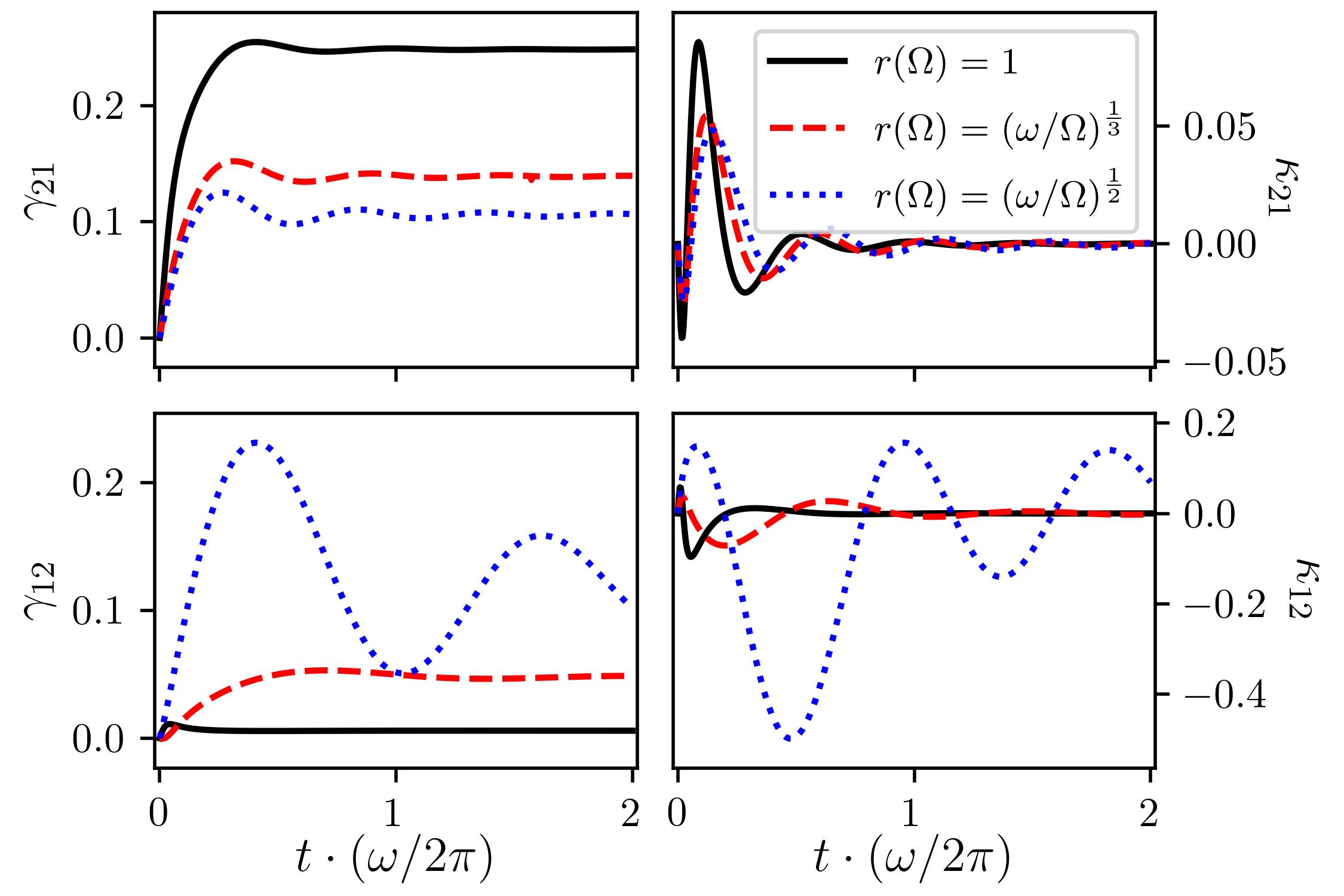}
\caption{The variations of decay rates with time when the environment is in a locally squeezed state with three different squeezing parameter dependencies $r(\Omega)=1$ (black solid lines), $r(\Omega)=(\omega/\Omega)^{1/3}$ (red dashed lines), and $r(\Omega)=(\omega/\Omega)^{1/2}$ (blue dotted lines).
}%
\label{fig:gammanm}%
\end{figure}
The system-reservoir interaction Hamiltonian now reads $H_{\rm int} \approx \sum_{\bf k}\frac{g_{\bf k}}{\sqrt{2^{n+m}}}(a^n b_{\bf k}^{\Dag m} +a^{\Dag n}b_{\bf k}^{m})$.
This can be written as $H_{\rm int} = \sum_{i=1,2} x_i X_i$, with $x_1 \equiv \sqrt{2^{-n}}a^n,~x_2 \equiv \sqrt{2^{-n}}a^{\Dag n},~X_1\equiv \sum_{\bf k}g_{\bf k}\sqrt{2^{-m}}b_{\bf k}^{\Dag m},~X_2\equiv \sum_{\bf k}g_{\bf k}\sqrt{2^{-m}}b_{\bf k}^{m}$.
Since the bath and system operators are no longer Hermitian, one thus, ends up with the following kernel
\begin{align}
\calK_2(t)\chi(t) &= -\sum_{j,k}\int_0^t\!ds\ \Big\{ \calC_{j,k}(t,s)\big[x_j(t),x_k(s)\rho(t)\big] +\text{H.c.} \Big\}\otimes R(t_0).
\label{rwakernel2}
\end{align}
The reservoir correlators are
\begin{subequations}
\begin{align}
\calC_{1,1}(t,t') &= \tr{\rm B}\{R(t_0) X_1(t)X_1(t')\} = \frac{1}{2^m}\sum_{\bf k}\sum_{\mathbf{k}'}g_{\bf k}g_{\mathbf{k}'}
\tr{\rm B}\Big\{
R(t_0)
b_{\bf k}^{\dag m} b_{\mathbf{k}'}^{\dag m} e^{im(\Omega_{\bf k}t +\Omega_{\mathbf{k}'}t')}
\Big\} = \int_0^\infty\! d\Omega\ J(\Omega) M_m^*(\Omega)e^{im\Omega(t +t')}, \\
\calC_{1,2}(t,t') &= \tr{\rm B}\{R(t_0) X_1(t)X_2(t')\} = \frac{1}{2^m}\sum_{\bf k}\sum_{\mathbf{k}'}g_{\bf k}g_{\mathbf{k}'}
\tr{\rm B}\Big\{
R(t_0)
b_{\bf k}^{\dag m} b_{\mathbf{k}'}^m e^{im(\Omega_{\bf k}t -\Omega_{\mathbf{k}'}t')}
\Big\} = \int_0^\infty\! d\Omega\ J(\Omega) N_m(\Omega)e^{im\Omega(t -t')}, \\
\calC_{2,1}(t,t') &= \tr{\rm B}\{R(t_0) X_2(t)X_1(t')\} = \frac{1}{2^m}\sum_{\bf k}\sum_{\mathbf{k}'}g_{\bf k}g_{\mathbf{k}'}
\tr{\rm B}\Big\{
R(t_0)
b_{\bf k}^m b_{\mathbf{k}'}^{\dag m} e^{-i m(\Omega_{\bf k}t -\Omega_{\mathbf{k}'}t')}
\Big\} = \int_0^\infty\! d\Omega\ J(\Omega) N_m'(\Omega)e^{-im\Omega(t -t')}, \\
\calC_{2,2}(t,t') &= \tr{\rm B}\{R(t_0) X_2(t)X_2(t')\} = \frac{1}{2^m}\sum_{\bf k}\sum_{\mathbf{k}'}g_{\bf k}g_{\mathbf{k}'}
\tr{\rm B}\Big\{
R(t_0)
b_{\bf k}^m b_{\mathbf{k}'}^m e^{-im(\Omega_{\bf k}t+\Omega_{\mathbf{k}'}t')} 
\Big\} = \int_0^\infty\! d\Omega\ J(\Omega) M_m(\Omega)e^{-im\Omega(t +t')}.
\end{align}
\end{subequations}
For a thermal reservoir one has $M_m=0$, $N_1=\Nbos$, $N_1'=\Nbos +1$, $N_2=\half\Nbos(\Nbos -1)$, and $N_2'=\half\big(\Nbos (\Nbos +3) +2\big)$.
Meanwhile, for a squeezed bath $M_1=-\half \sinh 2r\ e^{i\theta}$, $N_1=\sinh^2 r$, $N_1'=\sinh^2 r +1$, $M_2=\tfrac{3}{8} \sinh^2 2r\ e^{2i\theta}$, $N_2=\half\sinh^2 r\ (\sinh^2 r -1)$, $N_2'=\half\big(\sinh^2 r\ (\sinh^2 r +3) +2\big)$.
By plugging these back in Eq.~\eqref{rwakernel2} and moving back to the Schrodinger picture we arrive at the following RWA master equations in terms of creation and annihilation operators:
\begin{equation}
\dot\rho (t) =
-i[H_{\rm s}, \rho]
%-i\Delta_{n,m}(t) [a^\dag a, \rho]
+\gamma_{n,m}(t)\supD_{a^{\Dag n}}\rho
+\Gamma_{n,m}(t)\supD_{a^{n}}\rho
+\kappa_{n,m}(t)(\supD'_{a^{n}}\rho +\supD'_{a^{\Dag n}}\rho),
\label{masterrwa}
\end{equation}
%\dot\rho (t) =
%&-i[H_{\rm s}, \rho]
%-i\Delta_2(t) [a^\dag a, \rho]
%-\half i\Lambda_2(t) [(a^\dag a)^2, \rho]
%+\tfrac{1}{4}\gamma_2(t)\supD_{a^{\Dag 2}}\rho
%+\tfrac{1}{4}\big[\gamma_2(t) +\Gamma_2(t)\big]\supD_{a^{2}}\rho
%+\tfrac{1}{4}\kappa_2(t)(\supD'_{a^{2}}\rho +\supD'_{a^{\Dag 2}}\rho).
where we have assumed that $M_m^*=M_m$ ($\theta=0$ for a squeezed bath) and have introduced the dissipators $\supD_o\bullet \equiv 2o\bullet o^\dag -o^\dag o\bullet -\bullet o^\dag o$ and $\supD'_o\bullet \equiv 2o\bullet o -o o\bullet -\bullet o o$.
In the above equation we have also introduced the following parameters
\begin{subequations}
\begin{align}
%\lambda_n(t) &\equiv \int_0^\infty\!d\Omega J(\Omega)N(\Omega)\frac{\sin^2\big[\half(n\omega -\Omega)t\big]}{n\omega -\Omega}, \\
%\Lambda_n(t) &\equiv \int_0^\infty\!d\Omega J(\Omega)\frac{\sin^2\big[\half(n\omega -\Omega)t\big]}{n\omega -\Omega}, \\
\gamma_{n,m}(t) &\equiv \frac{1}{2^n}\int_0^\infty\!d\Omega J(\Omega)N_m(\Omega)\frac{\sin\big[(n\omega -m\Omega)t\big]}{n\omega -m\Omega}, \\
\Gamma_{n,m}(t) &\equiv \frac{1}{2^n}\int_0^\infty\!d\Omega J(\Omega)N_m'(\Omega)\frac{\sin\big[(n\omega -m\Omega)t\big]}{n\omega -m\Omega}, \\
\kappa_{n,m}(t) &\equiv \frac{1}{2^n}\int_0^\infty\!d\Omega J(\Omega)M_m(\Omega)\frac{\sin\big[(n\omega +m\Omega)t\big] -\sin(2m\Omega t)}{n\omega -m\Omega}.
\end{align}
\end{subequations}
The first term on the right hand side of the master equation gives the coherent evolution of the system with its renormalized frequency.
Note that in the two-photon interaction case ($n=2$) a Kerr nonlinearity is added to the Hamiltonian whose effect is negligibly small in the weak coupling regime that we are interested in.

%\end{widetext}
\twocolumngrid

%
%
%----------REFERENCES----------%
\bibliography{nMnC}

\end{document}